\documentclass[aps, prd, twocolumn, 10pt, superscriptaddress]{revtex4-2}
\usepackage[dvipsnames]{xcolor}
\usepackage{physics}
\usepackage{braket}
\usepackage{scalerel}
\usepackage{blindtext} 
\usepackage{epsfig, cancel}

\usepackage{latexsym}
\usepackage{natbib, comment}
\usepackage{mathrsfs,amsmath,amssymb,amsthm,amsfonts,tikz,graphicx,accents,hyperref,color}
\usepackage{url}
\usepackage{dcolumn}
\usepackage{multirow}
\usepackage{color}
\usepackage{cancel}
\usepackage{soul}
\usepackage[normalem]{ulem}
\usepackage{txfonts}
\usepackage{epsfig}
\usepackage{psfrag}
\usepackage{subfigure}
\hypersetup{colorlinks=true}
\usepackage{mathtools}
\usepackage{enumitem}
\usepackage{float}
\usepackage{caption,ragged2e}
\usetikzlibrary{decorations.markings,fpu}

\usetikzlibrary{calc}

\def\H0{{\text{H}\hspace*{-2.05mm}\text{H} 0\hspace*{-1.35mm}0\ }}

\usepackage[all]{xy}

\usepackage{slashed}
\usepackage{slashed,ccaption}
\usepackage{multirow}
\usetikzlibrary{calc} 

\usepackage{caption}

\usepackage{array}
\hypersetup{ linktoc=all,
    colorlinks, linkcolor={palatinateblue},
    citecolor={red}, urlcolor={amaranth} 
}

\graphicspath{{Images/}}

\renewcommand{\d}[1]{\ensuremath{\operatorname{d}\!{#1}}}

\DeclareSymbolFont{extraup}{U}{zavm}{m}{n}
\DeclareMathSymbol{\varheart}{\mathalpha}{extraup}{86}
\DeclareMathSymbol{\vardiamond}{\mathalpha}{extraup}{87}
\makeatletter
\renewcommand*{\@fnsymbol}[1]{\ensuremath{\ifcase#1\or \clubsuit \or \vardiamond \or \varheart\or
    \spadesuit\or \mathparagraph\or \|\or **\or \dagger\dagger
    \or \ddagger\ddagger \else\@ctrerr\fi}}
\makeatother

\definecolor{rosy}{RGB}{230,235,252}
\definecolor{myframetitle}{RGB}{90,89,170}
\definecolor{myblocktitle}{RGB}{140,185,249}
\definecolor{mytitle}{RGB}{10,80,26}

\definecolor{darkgreen}{RGB}{27,130,45}
\definecolor{darkblue}{rgb}{0,0,0.3}
\definecolor{darkred}{rgb}{0.7,0,0}

\definecolor{light gray}{RGB}{220,220,220}
\definecolor{dark purple}{RGB}{108,0,217}
\definecolor{pink}{RGB}{190,20,100}
\definecolor{orang}{RGB}{193,63,0}
\definecolor{green}{RGB}{11,98,17}
\definecolor{darkpink}{RGB}{153,0,76}
\definecolor{bluegreen}{RGB}{0,102,102}
\definecolor{greenlagan}{RGB}{0,102,0}
\definecolor{redgreen}{RGB}{102,102,0}
\definecolor{Redgreen}{RGB}{153,76,0}
\definecolor{vividviolet}{rgb}{0.62, 0.0, 1.0}
\definecolor{amaranth}{rgb}{0.9, 0.17, 0.31}
\definecolor{palatinateblue}{rgb}{0.15, 0.23, 0.89}
\definecolor{brightpink}{rgb}{1.0, 0.0, 0.5}
\definecolor{cornflowerblue}{rgb}{0.39, 0.58, 0.93}
\definecolor{deepcarminepink}{rgb}{0.94, 0.19, 0.22}
\definecolor{radicalred}{rgb}{1.0, 0.21, 0.37}
\newcommand\bc{\begin{center}}
\newcommand\ec{\end{center}}

\newcommand\ignore[1]{}
\usepackage[most]{tcolorbox}

\tcbset{highlight math style={left=02mm,right=02mm,top=02mm,bottom=02mm}} 
\usepackage{empheq}

\newcommand\inbox[1]{\tcbset{fonttitle=\scriptsize} \tcboxmath[colback=white,colframe=black!70]{#1}}

\begin{document}

\title{Dynamical Entropy Is a Noether Charge}
\author{V.~R.~Shajiee}\email{v.shajiee@ipm.ir}
\affiliation{School of Physics, Institute for Research in Fundamental Sciences (IPM), P.O.Box 19395-5531, Tehran, Iran}

\author{M.~M.~Sheikh-Jabbari}\email{jabbari@theory.ipm.ac.ir}
\affiliation{School of Physics, Institute for Research in Fundamental Sciences (IPM), P.O.Box 19395-5531, Tehran, Iran}

\author{V.~Taghiloo}\email{v.taghiloo@iasbs.ac.ir}
\affiliation{Department of Physics, Institute for Advanced Studies in Basic Sciences (IASBS), P.O.Box 45137-66731, Zanjan, Iran}
\affiliation{School of Physics, Institute for Research in Fundamental Sciences (IPM), P.O.Box 19395-5531, Tehran, Iran}

\begin{abstract}
{Black hole thermodynamics for generic dynamical, non-equilibrium regimes remains a fundamental challenge. We establish dynamical entropy as the Noether charge associated with a generic evolving null surface subject to Dirichlet boundary conditions. We specify the symmetry generator associated with the dynamical entropy, which is a null vector on the null surface, upon requiring physically motivated geometric conditions that yield a notion of ``dynamical zeroth law.''  We prove that this Noether charge density satisfies the second law of thermodynamics strictly at each instant in time, bypassing the teleological final conditions traditionally required by event horizons. Thus, we extend and generalize  the notion of  dynamical entropy introduced in \cite{Hollands:2024vbe}, in some different ways: 
We do not impose background stationarity; our dynamical entropy and the associated second law are local in time and work for generic dynamical gravitational systems.}

\end{abstract}
\maketitle


{Following the seminal discoveries of the 1970s \cite{Hawking:1971tu, Hawking:1971vc, Bekenstein:1973ur, Bardeen:1973gs, Hawking:1974sw, Hawking:1975vcx} and after 50 years of intensive work, the intimate relation between gravity and thermodynamics---best formulated in terms of black hole horizons---has become an established fact. Entropy occupies a central position in this framework, and its physical significance is encoded in the second law of thermodynamics. Accordingly, there should be a geometric observable associated with the entropy in systems involving black holes, or more generally in generic gravitating systems. The original Bekenstein-Hawking area law and Hawking's area theorem provide the first geometric definitions of entropy and its monotonic evolution.}

{The first mathematically rigorous and physically illuminating step in defining entropy for black holes was taken by Wald \cite{Wald:1993nt}, who demonstrated that for any diffeomorphism-invariant theory of gravity, the entropy of a stationary black hole is the Noether charge associated with the horizon-generating Killing vector field. This formulation, subsequently extended by Iyer and Wald \cite{Iyer:1994ys} and Jacobson, Kang, and Myers \cite{Jacobson:1993xs, Jacobson:1993vj}, establishes a robust geometric definition of entropy that reduces to the Bekenstein-Hawking area for General Relativity. However, the rigorous Noether charge derivation strictly applies only to stationary backgrounds in perfect equilibrium. By contrast, the true utility of entropy and the second law lies in dynamical, non-equilibrium processes.}

{Providing a mathematically rigorous notion of entropy and the second law beyond stationary black holes and for generic dynamical gravitational systems has remained a formidable challenge since the mid-1990s. Two influential programs have been pursued. First, Ashtekar and Krishnan \cite{Ashtekar:2002ag, Ashtekar:2003hk, Ashtekar:2004cn, Ashtekar:2025wnu} replaced the globally defined event horizon with a quasi-local, spacelike ``dynamical horizon.'' They derived an area-balance law whose matter and gravitational-wave fluxes are local and nonnegative, thereby providing dynamical first- and second-law statements in full nonlinear general relativity for the restricted class of horizons they consider.}

{More recently, Hollands, Wald, and Zhang (HWZ) \cite{Hollands:2024vbe} {(see also \cite{Visser:2024pwz, Rignon-Bret:2023fjq})} constructed a dynamical entropy for general diffeomorphism-invariant theories using covariant phase space formalism and a Wald-Zoupas-type \cite{Wald:1999wa} integrability prescription. The HWZ dynamical entropy formula applies to a dynamical horizon that is a null surface with a non-zero expansion and, at leading order, corrects Hawking's area-law by a term proportional to the expansion of the dynamical horizon. The HWZ dynamical entropy obeys first and second laws, only perturbatively around a stationary horizon. See  \cite{Ashtekar:2026jdz} for a discussion on limitations of the HWZ proposed dynamical entropy.

In this Letter, we bypass these limitations by introducing dynamical entropy as a Noether charge associated with an arbitrary generically expanding null surface. Our construction rests on the following two geometric and physical pillars:

\noindent (1) \textit{Dirichlet boundary conditions:} By supplementing the bulk gravitational action with the required null Gibbons-Hawking-York boundary term \cite{Parattu:2015gga, Lehner:2016vdi, Chandrasekaran:2021hxc}, we establish a well-posed variational principle through Dirichlet boundary conditions. This requirement isolates the boundary dynamics, allowing the null surface to behave as a thermodynamically closed system.

\noindent (2) \textit{Symmetry generator and dynamical zeroth law:} The evaluation of the Noether charge requires specifying the associated symmetry generator. We single out this symmetry generator by certain geometric (kinematical) requirements at the null surface. Crucially, one of them represents a notion of ``dynamical zeroth law'' that dynamically connects the surface gravity and the geometric expansion of the boundary.

Any qualified notion of entropy must satisfy the second law. We show that our dynamical entropy density does so upon using the Raychaudhuri equation, and once the matter field obeys the null energy condition (NEC). Our second law analysis is local in time and space and need not invoke teleological considerations \cite{Booth:2005qc}, unlike most of the second law discussions found in the literature \cite{Hawking:1971tu, Hawking:1973uf}. Unlike the Ashtekar--Krishnan construction, our boundary need not be a spacelike marginally trapped tube; unlike the HWZ work and discussions by Wall \cite{Wall:2015raa, Sarkar:2013swa}, no expansion about stationarity is invoked. Our analyses supply notions of dynamical zeroth law, a local-in-time entropy, and the second law in generic gravitating systems. Hence, our work complements Jacobson's seminal work \cite{Jacobson:1995ab} connecting gravity and thermodynamics. 


\bc\textbf{\large{Geometric Framework}}\label{sec:Geometry}\ec
Consider a $D$-dimensional Lorentzian manifold $(\mathcal{M}, g_{\mu\nu})$. Let ${\cal N}$ be a generic codimension-1 null hypersurface in ${\cal M}$, $l^\mu$ be the {future-oriented} null vector field generating ${\cal N}$; i.e., $l^\mu$ is tangent to the null geodesics in $\mathcal{N}$, $n^\mu$ be the null vector field transverse to ${\cal N}$. Moreover, we choose $l^\mu$ to be hypersurface orthogonal. The bulk spacetime metric $g_{\mu\nu}$ may be  decomposed as
\begin{equation}\label{metric-null-decomp}
    g_{\mu\nu} = -l_{\mu} n_{\nu} - l_{\nu} n_{\mu} + q_{\mu\nu} \, ,
\end{equation}
where $q_{\mu\nu}$ is the induced spatial metric residing on the $(D-2)$-dimensional transverse cross-sections of $\mathcal{N}$ and $l,n$ vector fields satisfy  orthogonality and normalization conditions,
\begin{equation}\label{null-normaliz}
    l \cdot l = 0\, , \quad n \cdot n = 0\, , \quad l \cdot n = -1\, ,\quad q_{\mu\nu} l^\nu = q_{\mu\nu} n^\nu = 0\,.
\end{equation}
The transverse spatial cross-sections of ${\cal N}$ will be denoted by $\Sigma$, $q_{\mu\nu}$ is the metric on $\Sigma$,  the antisymmetric binormal to it is
\begin{equation}\label{binormal-def}
    \epsilon_{\mu\nu} = l_{\mu} n_{\nu} - l_{\nu} n_{\mu} \, ,
\end{equation}
and is normalized as $\epsilon^{\mu\nu}\epsilon_{\mu\nu}=-2$.

The extrinsic geometry of $\mathcal{N}$ is governed by the covariant derivatives of $l_\mu, n_\mu$, which may be parameterized as,
\begin{align}
    \nabla_{\mu} l_{\nu} & = - \kappa_l n_{\mu} l_{\nu} - \kappa_n l_\mu l_\nu - n_{\mu} a_{\nu} + l_{\mu} \eta_\nu + \omega_{\mu} l_\nu + \Theta^l_{\mu\nu} \, , \label{eq:cov_l}\\
    \nabla_{\mu} n_\nu &= \kappa_l n_\mu n_\nu + \kappa_n l_\mu n_\nu - l_{\mu} \bar{a}_\nu - \omega_\mu n_\nu + n_\mu \bar{\eta}_\nu + \Theta^n_{\mu\nu} \, . \label{eq:cov_n}
\end{align}
Here, $\kappa_l$ and $\kappa_n$ denote the intrinsic non-affinity parameters (or surface gravities) associated with $l^\mu$ and $n^\mu$, respectively. The one-form $\omega_\mu$ characterizes the rotation (or twist) of the null surface, while $a_\mu$ and $\bar{a}_\mu$ act as transverse acceleration one-forms. The purely spatial tensors $\Theta^l_{\mu\nu}$ and $\Theta^n_{\mu\nu}$ encode the geometric deformation of the transverse cross-sections along their respective null directions.

Throughout this paper, equalities that hold strictly on the null boundary ${\cal N}$ will be denoted by the symbol $\mathrel{\hat{=}}$. The hypersurface orthogonality of $l^\mu$ then implies, 
\begin{equation}\label{hyper-ortho-l}
a_{\mu} \mathrel{\hat{=}} 0\, ,\qquad   \Theta^l_{\mu\nu} \mathrel{\hat{=}} \Theta^{l}_{\nu\mu}\, .
\end{equation}
We impose no such hypersurface orthogonality condition on  $n^\mu$, leaving $\Theta^n_{\mu\nu}$ as a generically non-symmetric tensor. Being a symmetric tensor on ${\cal N}$,  $\Theta^l_{\mu\nu}$ may be decomposed into traceless shear tensor $N_{\mu\nu}^l$ and the trace part expansion $\theta_l$:
\begin{equation}\label{devi-tens-decomp}
     \Theta^l_{\mu\nu} \mathrel{\hat{=}} N_{\mu\nu}^l + \frac{\theta_l}{D-2} q_{\mu\nu} \, .
\end{equation}
Utilizing the reparameterization freedom of the null generators $l_\mu$, we adopt an affine parameterization on the boundary:
\begin{equation}\label{affine-para-asum}
    \kappa_l \mathrel{\hat{=}} 0\, .
\end{equation}


Since $l^\mu$ is locally hypersurface orthogonal over $\mathcal{N}$, recalling the Frobenius theorem \cite{Wald:1984rg, Poisson}, $l_\mu$  can be parameterized as
\begin{equation}\label{normal-hyp-surf-ortho}
    l_\mu \mathrel{\hat{=}} - e^{-\Psi} \partial_\mu \Phi \, ,
\end{equation}
where the hypersurface $\mathcal{N}$ is located at $\Phi=0$. Assuming that the boundary ${\cal N}$ does not fluctuate, $\delta \Phi \mathrel{\hat{=}} 0$ we learn that, 
\begin{equation}\label{var-l}
    \delta l_\mu \mathrel{\hat{=}} -\delta \Psi\, l_\mu\, .
\end{equation}

\bc\textbf{\large{Variational Principle and Boundary Conditions}}\label{sec:Action-Variational}\ec
With the above equipment, we are now ready to derive dynamical entropy as the Noether charge.
To this end, we must first establish a mathematically well-posed variational principle for the gravitational theory with the null hypersurface $\mathcal{N}$ as its boundary and adopt physically appropriate boundary conditions by fixing the necessary boundary terms.

As argued, to discuss the entropy and the second law, we need to deal with a (gravitationally) closed system, which necessitates Dirichlet boundary conditions.  
For a spacetime region $\mathcal{M}$ equipped with a null boundary $\mathcal{N}$,  Dirichlet boundary conditions require the inclusion of a null analog of the Gibbons-Hawking-York (GHY) term \cite{Parattu:2015gga, Oliveri:2019gvm, Aghapour:2018icu, Chandrasekaran:2021hxc}. The total gravitational action is hence given by \footnote{{While $\sqrt{q}\theta_l = l\cdot \nabla \sqrt{q}$  is a total derivative along $l^\mu$, on ${\cal N}$, the boundary term in \eqref{action} cannot be dropped as a mere corner term, as the full null GHY term is $\sqrt{q}(\kappa_l + \theta_l)$, which is not a total derivative; only $\kappa_l + \theta_l$ is a valid scalar on $\mathcal{N}$ under reparameterizations of $l^\mu$. Thus, the $\theta_l$ term must be fully retained even when evaluated in the $\kappa_l = 0$ gauge.}},
\begin{equation}\label{action}
   \hspace{-.23 cm} S = \frac{1}{16\pi G} \int_{\mathcal{M}} \d{}^{D}x \sqrt{-g} (R-2\Lambda) + \frac{1}{8\pi G} \int_{\mathcal{N}} \d{}^{D-1}x\sqrt{q}\, \theta_l.
\end{equation}
Note that the null GHY boundary term is $\sqrt{q}(\kappa_l + \theta_l)$, and in our adapted affine parameterization \eqref{affine-para-asum} yields the one in \eqref{action}.

Variation of the total action \eqref{action} gives the standard bulk equations of motion, along with a boundary-variation term $\boldsymbol{\Theta}(\mathcal{N})$. It is well known that $\boldsymbol{\Theta}(\mathcal{N})$ is subject to the so-called $W$ and $Y$ freedoms (or ambiguities) \cite{Iyer:1994ys, Wald:1999wa}. They
may be (partially) fixed by imposing a well-defined action principle with appropriate boundary conditions and requiring $W$ and $Y$ to be covariant with respect to the null boundary geometry. The former implies that upon imposing the boundary conditions $\boldsymbol{\Theta}(\mathcal{N})$ should become an integral over a codimension-2 surface, constant time slice on ${\cal N}$ \cite{Harlow:2019yfa, Shi:2020csw, Sheikh-Jabbari:2025kjd}.

The null GHY boundary term is compatible with fixing the  Dirichlet boundary conditions, defined through \eqref{var-l} and $\delta q_{\mu\nu} \mathrel{\hat{=}} 0\, , \delta l^{\mu} \mathrel{\hat{=}} 0$ \cite{Parattu:2015gga, Lehner:2016vdi, Chandrasekaran:2021hxc}. This boundary term appropriately fixes $W$ and $Y$ freedoms up to codimension-2 corner terms, yielding,
\begin{align}\label{eq:Theta_initial}
    \boldsymbol{\Theta}(\mathcal{N}) := \int_{\mathcal{N}}\d{}x_{\mu} \boldsymbol{\Theta}^{\mu} &= \frac{1}{16\pi G} \int_{\mathcal{N}} \d{}^{D-1}x \sqrt{q} l_{\mu} (\nabla_{\nu}h^{\mu\nu} - \nabla^{\mu}h) \nonumber \\
    & + \frac{1}{8\pi G} \int_{\mathcal{N}} \d{}^{D-1}x\, \delta (\sqrt{q}\, \theta_l) \, ,
\end{align}
with $h_{\mu\nu}:= \delta g_{\mu\nu}$ denoting an arbitrary variation of the bulk metric, leads to a well-defined variational principle. Once $\boldsymbol{\Theta}(\mathcal{N})$ is specified, e.g. as in \eqref{eq:Theta_initial}, we have all the needed information for an unambiguous computation of the Noether charge.

\bc\textbf{\large{Dynamical Entropy Symmetry Generators}}\label{sec:Symmetry-Generator}\ec

After fixing the boundary conditions and the boundary variation term  $\boldsymbol{\Theta}(\mathcal{N})$, the next step in computing a Noether charge is to specify the corresponding symmetry generator. Here, we  single out this future-oriented symmetry generator $\xi$ via the following kinematic constraints evaluated strictly on the null hypersurface $\mathcal{N}$:
\begin{subequations}\label{xi-properties}
    \begin{align}
        \xi \cdot \xi & \mathrel{\hat{=}} 0 \, , \label{xi-prop-a} \\
        \xi \cdot \nabla \xi & \mathrel{\hat{=}} \kappa\, \xi \, , \label{xi-prop-b} \\
       \epsilon^{\mu\nu}\nabla_{\mu} \xi_{\nu} &  \mathrel{\hat{=}} -2 \kappa \, , \label{xi-prop-c} \\
         \nabla\cdot \left(\frac{\xi}{\kappa}\right) & \mathrel{\hat{=}}  0\, .\label{g-zero-law}
    \end{align}
\end{subequations}
The above is a generalization/extension of properties of a Killing horizon-generating vector field $\xi$ with surface gravity $\kappa$, to a generic null surface ${\cal N}$, which is dynamical.
We pause to discuss the physical meaning of the above defining relations:
\begin{enumerate}[label=\Roman*., leftmargin=-0.5mm, itemsep=0pt]
   \item Eq.~\eqref{xi-prop-a} guarantees that the flow of $\xi$ remains tangent to the null boundary ${\cal N}$;
    \item Eq.~\eqref{xi-prop-b} identifies $\xi$ as a geodesic generator with non-affinity parameter (local surface gravity) $\kappa$. $\kappa/(2\pi)$ plays the role of \textit{dynamical temperature};
    \item Eq.~\eqref{xi-prop-c} relates $\xi$ to the generator of local boosts in the $2d$ plane spanned by $l_\mu, n_\mu$, see \cite{Shajiee:2025cxl} for more discussions. This condition ensures that the non-affinity parameter $\kappa$ corresponds to the boost weight of the generator;
    \item The zeroth law of thermodynamics defines the notion of thermodynamic equilibrium and when thermodynamic description is applicable. In other words, a thermodynamic description of generically evolving systems is meaningful only when one provides a notion of zeroth law. As we will show below, \eqref{g-zero-law} yields a notion of \textit{generalized zeroth law} which guarantees applicability of thermodynamic description to generic dynamical boundaries that are arbitrarily far from stationarity. 
    
\end{enumerate}

The next task is to construct the most general solutions to \eqref{xi-properties}  in terms of the geometric null basis $(l^\mu, n^\mu)$ introduced earlier. To do so, we introduce the dual vector field $\chi_\mu$,
\begin{equation}\label{N-properties}
    \chi\cdot \chi \mathrel{\hat{=}} 0 \, , \qquad \xi \cdot \chi \mathrel{\hat{=}} -1 \, .
\end{equation}
Eqs. \eqref{xi-properties} define the vector field $\xi$, locally in the neighborhood of the null boundary ${\cal N}$,  up to foliation-preserving diffeomorphisms on ${\cal N}$, which always allows us to set $q_{\mu\nu} \xi^\nu=0$. Therefore, one can always  expand $\xi$ as 
\begin{align}\label{def-xi}
     \xi^\mu & = \kappa(\mathscr{B} l^\mu  - \mathscr{A} n^\mu) \, , \\
     \chi^{\mu} & = \kappa^{-1}(\mathscr{B}^{-1} n^{\mu} - \mathscr{D} l^{\mu} )\, ,
\end{align}
where $\mathscr{B}$, $\mathscr{A}$, and $\mathscr{D}$ are generic scalar functions and $\xi$ being future-oriented implies $\mathscr{B} \ge 0$. Substituting these  into \eqref{xi-prop-a} and the first relation in \eqref{N-properties}, leads to
\begin{equation}\label{U-u-conditions}
    \mathscr{A} \mathrel{\hat{=}} 0 \, , \qquad \mathscr{D} \mathrel{\hat{=}} 0 \, .
\end{equation}
That is, on ${\cal N}$, $\xi$ is along the geometric null normal $l^\mu$, while $\chi^\mu$ is proportional to the auxiliary normal $n^\mu$. However, away from ${\cal N}$, $\xi, \chi$ move away from $l, n$ respectively. 

Imposing \eqref{xi-prop-b} and \eqref{xi-prop-c}, using \eqref{eq:cov_l}, fixes the gradients of the scalar functions evaluated on the boundary as
\begin{align}\label{cov-BB}
    \nabla_{\mu} \mathscr{A} \mathrel{\hat{=}} - l_{\mu} \, ,\qquad 
    l\cdot \nabla (\kappa\mathscr{B})  \mathrel{\hat{=}} \kappa\, .
\end{align}
Finally,  recalling \eqref{eq:cov_l} and \eqref{eq:cov_n}, \eqref{g-zero-law} yields, 
\begin{equation}\label{gen-zeroth-law}
  \inbox{  l \cdot \nabla \ln\mathscr{K} \mathrel{\hat{=}} 0 \,, \quad \mathscr{K}:= \frac{\kappa}{\sqrt{q}}\, ,}
\end{equation}
implying constancy of the local ``temperature density'' $\mathscr{K}$ along $l^\mu$; justifying the name the generalized (local) zeroth law.

\bc\textbf{\large{Dynamical Entropy Is a Noether Charge}}\label{subsec:Noether}\ec
With the symmetry generator specified, we can use the standard, textbook analysis to calculate the dynamical entropy as the Noether charge corresponding to $\xi$, e.g., see \cite{Wald:1993nt, Grumiller:2020vvv}.  The Noether current associated with $\xi$ is: 
\begin{equation}
    \mathbf{J}_\xi^{\mu} = \boldsymbol{\Theta}^{\mu}[\delta_\xi g]-\xi^{\mu} \text{L}\, , \quad \mathbf{J}^{\mu}_\xi = \partial_{\nu} \text{Q}_\xi^{\mu\nu}\, ,
\end{equation}
where L is the total Lagrangian given in \eqref{action}, which includes the GHY boundary term and $\boldsymbol{\Theta}^{\mu}$ is given by \eqref{eq:Theta_initial}. The second equality is valid on-shell and is a consequence of the fact that $\nabla\cdot \mathbf{J}_\xi=0$ on-shell. Thus, the Noether surface charge, which should be evaluated on-shell, is obtained as:
\begin{equation}\label{Noe-formula}
\begin{split}
\text{Q}^{\text{\tiny{N}}}_\xi := & \oint_{\Sigma} \d{}x_{\mu\nu} \ \text{Q}^{\mu\nu}_\xi \\
= & - \frac{1}{16\pi G} \oint_{\Sigma} \d{}^{D-2}x\,\sqrt{q}\, \epsilon_{\mu\nu}\, \left(\nabla^{[\mu}\xi^{\nu]} + 2\xi^{[\mu}\mathcal{W}^{\nu]}\right) \, ,
\end{split}
\end{equation}
where $\mathcal{W}^{\mu} = - n^{\mu}\,  \theta_l$ is the contribution from the null GHY boundary Lagrangian established in \eqref{action}. The freedoms/ambiguities in the symplectic potential $\boldsymbol\Theta({\cal N})$ yield ambiguities in the Noether charges \cite{Wald:1993nt, Iyer:1994ys}. As discussed above,  we have fixed these ambiguities by specifying  $\boldsymbol\Theta({\cal N})$ as in \eqref{eq:Theta_initial}.

Evaluating the Noether charge for our dynamically adapted generator $\xi$, utilizing  \eqref{def-xi} and \eqref{cov-BB}, yields
\begin{equation}\label{Dyn-Entropy-Noe}
  \text{Q}^{\text{\tiny{N}}}_\xi = \frac{1}{8\pi G} \oint_{\Sigma} \d{}^{D-2}x\, \sqrt{q}\, \kappa \left( 1 - \mathscr{B} \theta_l \right) \, . 
\end{equation}
In standard gravitational thermodynamics, the total Noether charge evaluated on a horizon identifies the thermodynamic heat, $\text{Q}^{\text{\tiny{N}}}_\xi = \oint_{\Sigma} \frac{\kappa}{2\pi} \mathcal{S} \d{}^{D-2}x$ \cite{Wald:1993nt, Jacobson:1995ab, Hajian:2015xlp, Adami:2021kvx}. Recalling the Clausius relation, one may factor out the local temperature ${\kappa}/({2\pi})$ from \eqref{Dyn-Entropy-Noe}, to obtain \textit{dynamical entropy density} \footnote{Eqs.~\eqref{xi-properties} and the dynamical entropy \eqref{eq:charge_density} are invariant under a global rescaling $(\xi, \kappa) \to (\alpha \xi, \alpha \kappa)$ for a constant $\alpha$. This freedom reflects the choice of a global time unit.}:
\begin{equation}\label{eq:charge_density}
    \inbox{\mathcal{S} \mathrel{\hat{=}} \frac{\sqrt{q}}{4 G} \left( 1 - \mathscr{B} \theta_l \right) \, .}
\end{equation}
This expression precisely coincides with the HWZ \cite{Hollands:2024vbe} dynamical entropy \footnote{The dynamical term in \eqref{eq:charge_density} has previously appeared in the literature \cite{Mishra:2017sqs, Adami:2021nnf, Adami:2021kvx}. However, it has not been interpreted as contributing to the entropy.}. One can show that this is the integrable part of the surface charge variation derived via the covariant phase space formalism \cite{In-progress}. Therefore, upon requiring Dirichlet boundary conditions and a well-defined variational principle, the HWZ dynamical entropy can be obtained through the standard Noether charge of the symmetry generator $\xi$ for generic gravitational systems in arbitrarily far-from-stationarity cases.

\bc\textbf{\large{Investigating the Second Law }}\label{sec:Second-Law}\ec
Any valid entropy should satisfy the second law: it should have monotonic growth over time. To investigate the validity of the second law, we study the evolution of the dynamical entropy density $\mathcal{S}$ \eqref{eq:charge_density} along the null boundary $\mathcal{N}$; i.e., we evaluate the Lie derivative of  $\mathcal{S}$  along the null vector $l^\mu$. Recalling \eqref{cov-BB} and that  $l \cdot \nabla \sqrt{q} = \sqrt{q} \theta_l$, we obtain
\begin{equation}\label{eq:entropy_derivative_intermediate}
    l \cdot \nabla \mathcal{S} \mathrel{\hat{=}} \frac{\sqrt{q}}{4 G} \mathscr{B} \Big[ - l \cdot \nabla \theta_l + \theta_l (l \cdot \nabla \ln\kappa  - \theta_l)\Big] \, .
\end{equation}
As also pointed out in \cite{Hollands:2024vbe}, cancellation of the linear $\theta_l$ terms ($\theta_l - \theta_l = 0$) is a crucial structural feature of the dynamical entropy, ensuring that leading-order geometric area fluctuations do not artificially drive the entropy variation negative. Finally, upon the generalized zeroth law \eqref{gen-zeroth-law}, we remain with a  remarkably simple evolution equation:
\begin{equation}
    l \cdot \nabla \mathcal{S} \mathrel{\hat{=}} -\frac{\sqrt{q}}{4 G} \mathscr{B} (l \cdot \nabla \theta_l) \, .
\end{equation}

To assess the sign of $l \cdot \nabla \mathcal{S}$ we recall that the evolution of the expansion scalar $\theta_l$ along $l^{\mu}$ null rays is governed by the well-known purely geometric Raychaudhuri equation \cite{Hawking:1973uf}, which, recalling \eqref{hyper-ortho-l} and \eqref{affine-para-asum}, takes the form
\begin{equation}\label{Raychaudhuri}
    l \cdot \nabla \theta_l + \frac{\theta_l^2}{D-2} + N_{l}^2 + R_{ll} \mathrel{\hat{=}} 0 \, .
\end{equation}
Here, $N_{l}^2 := N^{\mu\nu}_{l}N_{\mu\nu}^{l} \ge 0$ is non-negative. The term $R_{ll} = l^{\mu}l^{\nu}R_{\mu\nu}$ is the component of the bulk Ricci tensor along $l^\mu$ and, through Einstein equations, represents the local matter flux crossing the boundary. Note that the cosmological constant in \eqref{action} does not contribute to $R_{ll}$. Therefore,
\begin{equation}\label{eq:second_law_final}
    \inbox{ l \cdot \nabla \mathcal{S} \mathrel{\hat{=}} \frac{\sqrt{q}}{4 G} \mathscr{B} \left( N_l^2 + \frac{\theta_l^2}{D-2} + R_{ll}   \right)\, .}
\end{equation}
As mentioned (cf. discussion below \eqref{def-xi}) $\mathscr{B}\geq 0$  and the $N_l^2$ and ${\theta_l^2}$ terms are non-negative by definition. So, if $R_{ll} \ge 0$, which implies NEC on the bulk matter fields through Einstein field equations, it guarantees the second law,
\begin{equation}
    l \cdot \nabla \mathcal{S} \ge 0 \, ,
\end{equation}
for the dynamical entropy density \eqref{eq:charge_density},  the Noether charge associated with vector field $\xi$, if the matter field satisfies NEC and if the dynamical zeroth law \eqref{gen-zeroth-law} holds. Note  that these are all locally defined around a generic null surface ${\cal N}$.

\bc\textbf{\large{Discussion}}\label{sec:Conclusion}\ec
More than thirty years ago, Wald published a seminal paper entitled ``Black hole entropy is the Noether charge'' \cite{Wald:1993nt}, and Wald and collaborators \cite{Hollands:2024vbe} recently provided an expression for the entropy of nonstationary (dynamical) black holes. Inspired by these works, we constructed a notion of dynamical entropy which can be obtained as a Noether charge, which has the same expression as the HWZ entropy but works for a generic dynamical null surface (an extension of notion of horizon) and along the way we discussed a notion of dynamical zeroth law and proved that this dynamical entropy satisfies the second law, if the matter content satisfies NEC.

As discussed, our notion of the entropy goes beyond black holes and applies to any spacetime region bounded by a null boundary. Therefore, our results and analyses should have implications for wherever gravity is relevant, e.g., in cosmology. Moreover, our analysis here indicates that the thermodynamic description of gravity is a quasi-local one in the sense of Brown-York \cite{Brown:1992br} and that it involves surface charge and applies to an arbitrary codimension-1 surface positioned in an arbitrary place in spacetime. These statements and results resonate with those of Jacobson \cite{Jacobson:1995ab}, see also \cite{Adami:2021kvx}, and may teach us more about the nature of gravity.

Gravity on spacetimes with null boundaries and associated surface/boundary charges has been studied in several papers, e.g., \cite{Donnay:2015abr, Donnay:2016ejv, Hopfmuller:2018fni, Chandrasekaran:2018aop, Grumiller:2019fmp,  Adami:2020amw, Grumiller:2020vvv, Adami:2020ugu} and in particular in \cite{Adami:2021nnf, Adami:2021kvx}. It is desirable to connect the analysis in these works, which are based on covariant phase space formalism, to our analysis here, which is based on Noether charges \cite{In-progress}. This will hence make a direct connection between our discussions and those in \cite{Hollands:2024vbe}.

Finally, the notion of entropy and the second law is not limited to Einstein gravity, and it is expected to work for any higher-curvature gravity theory, see e.g. \cite{Jacobson:1995uq, Sarkar:2013swa, Wall:2015raa, Hollands:2022fkn, Davies:2023qaa, Bhattacharyya:2021jhr}. Both the Noether charge method and the covariant phase space formalism constructions are applicable to higher curvature theories. It is instructive to analyze the results here beyond Einstein gravity theories.

\vspace{.5 cm}
The work of VRSh acknowledges Iran National Science Foundation (INSF) grant No. 4031681. MMSh-J would like to thank the INSF research chair grant No. 4045163.

\bibliographystyle{fullsort.bst}
\bibliography{reference}
\end{document}